\documentclass[nofootinbib,twocolumn,aps,prl,superscriptaddress,longbibliography]{revtex4-2}

\usepackage[utf8]{inputenc}
\usepackage[T1]{fontenc}
\usepackage{bm}
\usepackage{amsmath,amssymb}
\usepackage{graphicx}
\usepackage{xcolor}
\usepackage{hyperref}
\usepackage{natbib}

\graphicspath{{Figures/}}

\begin{document}

\title{Reaching diffraction-limited localization with coherent PTAs}

\author{Anna C. Tsai}
\email{anna.tsai@mail.utoronto.ca}
\affiliation{Canadian Institute for Theoretical Astrophysics, 60 St. George Street, Toronto, ON M5S 3H8, Canada}
\affiliation{Department of Physics, University of Toronto, 60 St. George Street, Toronto, ON M5S 1A7, Canada}
\affiliation{Dunlap Institute for Astronomy \& Astrophysics, University of Toronto, AB 120-50 St. George Street, Toronto, ON M5S 3H4, Canada}

\author{Dylan L. Jow}
\email{dylanjow@stanford.edu}
\affiliation{Kavli Institute for Particle Astrophysics and Cosmology, Stanford University, 452 Lomita Mall, Stanford, CA 94305-4085, USA}

\author{Ue-Li Pen}
\affiliation{Institute of Astronomy and Astrophysics, Academia Sinica, Astronomy-Mathematics Building, No. 1, Section 4, Roosevelt Road, Taipei 10617, Taiwan}
\affiliation{Canadian Institute for Theoretical Astrophysics, 60 St. George Street, Toronto, ON M5S 3H8, Canada}
\affiliation{Department of Physics, University of Toronto, 60 St. George Street, Toronto, ON M5S 1A7, Canada}
\affiliation{Dunlap Institute for Astronomy \& Astrophysics, University of Toronto, AB 120-50 St. George Street, Toronto, ON M5S 3H4, Canada}
\affiliation{Perimeter Institute for Theoretical Physics, 31 Caroline St. North, Waterloo, ON N2L 2Y5, Canada}
\affiliation{Canadian Institute for Advanced Research, CIFAR Program in Gravitation and Cosmology, MaRS Centre, West Tower, 661 University Ave., Suite 505, Toronto, ON M5G 1M1, Canada}

\begin{abstract}
Current pulsar timing array (PTA) analyses are phase incoherent and thus to not take full advantage of pulsar distance information, thereby missing out on improved angular resolution and on a potential factor-of-two gain in detection sensitivity for individual gravitational-wave (GW) sources. In this work, we investigate the impact of precise pulsar distance measurements on angular resolution as an extension to previous work measuring the angular resolution of a dense, isotropic PTA [Jow et al., 2025]. We present a coherent map-making technique that utilizes precise pulsar distance measurements to reach a diffraction-limited resolution of an individual source: $\delta \theta_{\mathrm{diff}} \times (1/\mathrm{SNR}) \approx 1~\mathrm{arcmin}$, where $\delta \theta_{\mathrm{diff}} = \lambda_{\mathrm{GW}}/r$ is the diffractive angle and SNR refers to the detection strength of the source. With this level of angular resolution, identifying an EM counterpart may become feasible, enabling multi-messenger follow-up. We show that for $\rm SNR=10$, which may be the current sensitivity level using a coherent analysis, the diffraction limit is reached with approximately 10 equidistant pulsars with distances of about 300 parsecs. Moreover, angular resolution scales sharply with the number of known pulsar distances as $\sim (1/\mathrm{SNR})^{N_{\mathrm{dist}}/2}$. Thus, each additional pulsar with high signal-to-noise timing and precise distance measurement can improve PTA resolution by an order of magnitude. The distance to the best-timed millisecond pulsar (PSR J0437$-$4715) is already constrained to sub-parsec levels. We argue, therefore, that a coherent analysis of PTA data, fully incorporating pulsar distance information, is timely.
\end{abstract}

\maketitle


\section{Introduction}

Constructing high-resolution maps of the nanohertz gravitational wave (GW) sky will inform inferences of the origin of the gravitational
wave background (GWB). Standard pulsar timing array (PTA) analyses have focused on measuring the amplitude of the signal treated as an
isotropic stochastic background. While this method has provided recent evidence for the detection of a nanohertz gravitational-wave
background (GWB) by multiple PTA collaborations~\citep{Nanograv15yr,IPTA,EPTA,CPTA,PPTA}, an astrophysically-sourced GWB (e.g. from inspiraling supermassive blackhole binaries) is neither stochastic nor isotropic.  A more
informative approach is to construct maps of the GW sky \citep{Grunthal, Taylor, Ali-Haimoud, Cornish}. 

Distinguishing between GWB origins is particularly salient in light of a possible tension between the GWB amplitude measured and the GWB amplitude expected from estimated source populations~\citep{Sato-Polito}. One possible resolution to this tension is the presence of a handful ($\sim 1$--$10$) nearby (within $100~{\rm Mpc}$) bright sources in some frequency bins that may be considerably more massive than previously expected~\citep{Sato-Polito}. Resolved maps of the GW sky can potentially identify these bright sources, enabling multi-messenger follow up. In particular, confirming an association between quasars showing periodic optical variability \citep[e.g.][]{Charisi, Graham} and nanohertz gravitational wave sources would open up a powerful electromagnetic counterpart (EM) channel \citep{Petrov}. However, maps with high resolution will be needed to sufficiently localize the GW sources in order to make these associations. Individual source localization is crucial to proposed measurements of cosmic expansion using nanohertz GWs. Such measurements can be obtained either through a standard-sirens approach—combining an EM-counterpart redshift with a GW-inferred luminosity distance from the small frequency shift between the Earth and pulsar terms \citep{D'Orazio}—or via diffractive lensing by edge-on galaxies \citep{Jow2025}. In addition to compact astrophysical sources, the background may include a cosmological component (arising from e.g. cosmic strings,
primordial GWs from inflation, phase transitions, or acoustic shocks~\citep{PenTurok}). Identifying and removing astrophysical sources will be necessary for detecting a cosmological component.  

In principle, a PTA's maximum angular resolution is set by the diffractive angle, \citep{BoylePen} 
\begin{equation}
    \delta \theta_{\rm diff}\sim \frac{\lambda_{\rm GW}}{r}
\end{equation}
where $r$ is the pulsar distance and $\lambda_{\rm GW}$ is the gravitational wavelength. We consider a PTA with an SNR of 10 (which we will later motivate as the current signal strength seen by PTAs if the signal were primarily due to a single bright source). Here, SNR refers to the detection level of the source in the matched-filter map (not the SNR of the timing residuals). For a reference pulsar distance of $\sim 300~{\rm pc}$, and a gravitational wavelength of $1~{\rm pc}$, the diffraction limited localization is approximately $\delta \theta_{\mathrm{diff}} \times (1/\mathrm{SNR}) = 1$ arcmin,
which corresponds to a fraction of the sky of roughly $10^{-8}$. As many pulsars are a kiloparsec or more away, the actual resolution may be higher. Estimates of the total number of sources emitting GWs in the
$10^{-7}$--$10^{-9}~{\rm Hz}$ range are at most $\sim
10^{6}$~\citep{Sesana}. Thus, it is
unlikely that an astrophysical background will be in the
so-called confusion limit of a diffraction limited coherent PTA. However, achieving this hypothetical resolution limit will require a full, phase informed inclusion of pulsar distance information in PTA analyses.

In ~\citet{Jow}, we promote the
construction of matched-filter maps of the GW sky in order to search
for continuous sources and distinguish between astrophysical and
cosmological origins of the GWB, showing that matched-filter maps
are information-preserving and model-agnostic transformations of the
data. We constrained the number of resolvable GW
sources for a uniform PTA using matched-filter map-making techniques
using only the so-called ``Earth-term'', which does not require known pulsar distances. In this work, we continue our
investigation into PTA angular resolution by considering the effects
of the pulsar-distance dependent term in the timing residual: the ``pulsar-term''.  The
Earth term is the change in the time of arrival (TOA) due to the
Earth's time dilation as a GW passes through the PTA. The pulsar term is the
contribution from the time dilation at the pulsar. Developing the ability to
use PTAs as standard interferometric arrays involves both terms, since
it is the Earth-pulsar pair that forms the antenna. Not only does using pulsar distance information to measure GW phase enable diffraction-limited resolution, but it also accounts for approximately half of the signal strength, potentially boosting the detection significance by twofold. However, leveraging the
pulsar term requires knowing the distances to the pulsars, which for
the most part, are measured with uncertainties much higher than the
gravitational wavelength of interest.  Ref.~\cite{Lee} places constraints on the observable GW amplitude for a single source using a coherent analysis. Ref.~\cite{Kato} and Ref.~\cite{Kato2025}
demonstrate how precisely constrained distances improve resolution for
specific PTA configurations.  Our goal is to answer the question: how
many constrained pulsar distances are needed to reach the
diffraction-limited resolution?

In this paper, we modify the matched-filter map making technique to
include the pulsar term.  We construct coherent maps for increasing
numbers of approximately uniformly distributed pulsars across the sky
with precisely constrained pulsar distances in addition to a dense
array of pulsars with unknown distances.  We estimate the angular
resolution for a single GW source by measuring the area above an
iso-contour on a single source map. We introduce a scaling relation, derived in the flat sky limit local to the GW source, demonstrating how angular resolution improves as a function of the number of precisely constrained pulsar distances. We use full-sky simulations with very short pulsar distances to produce large interference structures, which are numerically simple to resolve, to demonstrate the coherent matched-filter map making method and to support the scaling relation. We estimate we will reach
diffraction-limited angular resolution with a modest number of roughly
$10$ well-constrained pulsar distances at an aggregate SNR level of 10.
Additionally, due to the exponential form of the scaling relation, we show that the addition of one or two precisely
constrained pulsar distances to the PTA sample can increase the
point-source resolution by an order of magnitude. We, therefore, argue
that constraining pulsar distances is a timely project that could have a high impact on PTA science in the near term. 


\section{Constructing coherent maps of the GW sky}


The main observable from a PTA is the Fourier transform of the pulsar timing residual, which for a pulsar at position ${\bm r}$ induced by a monochromatic GW source propagating in the $\hat{n}$ direction is given by:

\begin{align}
    &\tilde{\tau}^{R,L}(\boldsymbol{r}) = -\frac{h}{4 \pi f_0} \zeta^{R,L}(\boldsymbol{r};\hat{n},t_0), \\
    \begin{split}
    &\zeta^{R,L}(\boldsymbol{r};\hat{n},t_0) = e^{2 \pi i f_0 t_0} \,
    \times \\
    &\frac{\left(\cos\theta \sin \theta_{\rm gw} - \cos \theta_{\rm gw} \cos \left( \phi - \phi_{\rm gw} \right) \sin \theta \pm i \sin \theta \sin \left( \phi - \phi_{\rm gw}\right) \right)^2}{\cos\theta \cos \theta_{\rm gw} + \cos\left(\phi - \phi_{\rm gw} \right) \sin\theta \sin \theta_{\rm gw} - 1} \\
    & \times \left( 1- \mathcal{P}(f;\boldsymbol{r})\right),
    \end{split}
\end{align}
where $\mathcal{P}(f;\boldsymbol{r}) = e^{2 \pi i f r (1 + \hat{n} \cdot \hat{r})}$, $\hat{n} = (1,\theta_{gw},\phi_{gw})$ is the unit vector pointing in the direction of GW propagation, and $\boldsymbol{r} = (r,\theta,\phi)$ is the position of the pulsar in relation to the Earth (in spherical coordinates).  R and L indicate that the source is right-circularly or left-circularly polarized. Arbitrary source polarizations can be achieved by a linear combination of these two modes. The scale of phase oscillations in the pulsar term depends only on the ratio of pulsar distance to GW wavelength; therefore, we introduce a dimensionless parameter $\nu \equiv r/\lambda_{\rm GW}$ and $\mathcal{P}(f;\boldsymbol{r})$ becomes $\mathcal{P}(f;\boldsymbol{r}) = e^{i 2 \pi \nu (1 + \hat{n} \cdot \hat{r})}$.

\begin{figure}[htbp]
    \centering
    \includegraphics[width=1\linewidth]{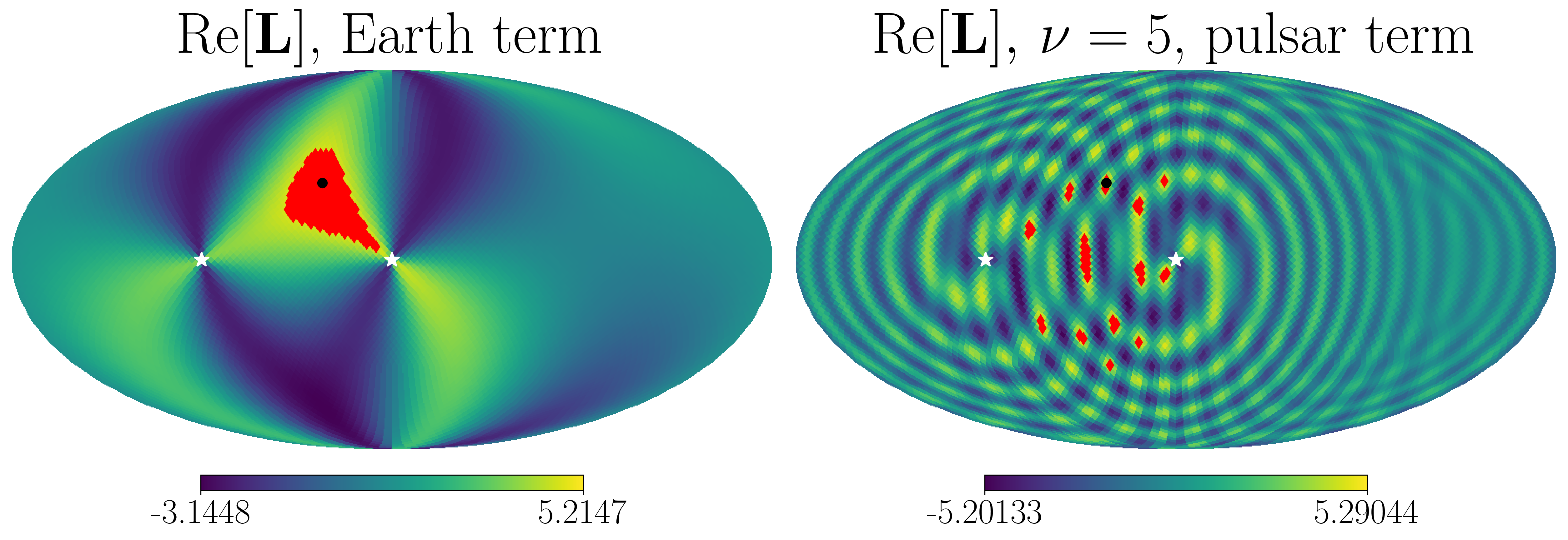}
    \caption{The real part of $\bf L$ (left circular polarization) for an Earth term only matched-filter map (left) and a pulsar term only map (right).  Both maps are in response to the same left-handed GW located at the black dot.  The red points mark regions of the matched-filter map where pixel values are above $90 \%$ of the maximum pixel (located at the GW position). The array consists of two pulsars with identical distances and $\nu=5$.  This figure demonstrates the fringe interference effects from the pulsar term that generates isolated islands of intensity that allow for GW localization far beyond the limit of an Earth-term only analysis.}
    \label{fig:nu30}
\end{figure}

We introduce the aperture matrices:
\begin{align}
    A^L_{ij} (r)&= \zeta^L(\boldsymbol{r}_i;\hat{n}_j,\hat{n}_p=+\hat{z}), \\
    A^R_{ij} (r)&= \zeta^R(\boldsymbol{r}_i;\hat{n}_j,\hat{n}_p=-\hat{z}),
\end{align}
which depend on the full pulsar position vector.  Following our previous work, we construct the left and right matched-filter maps, which estimate the GW distribution across the sky \citep{Jow}:
\begin{align}
    {\bf L} &= (A^L (r))^\dagger {\bf d},\\
    {\bf R} &= (A^R (r))^\dagger {\bf d},
\end{align}
where ${\bf d} = [\tilde{\tau}_1,...,\tilde{\tau}_N]$ is the time Fourier transform of the pulsar timing residuals for each pulsar in the array.  Here, as in Ref.~\cite{Jow}, we assume that there is no correlation between the pulsar noise in constructing the matched filter maps. In Ref.~\cite{Jow}, we assume that the pulsar distances are effectively unknown.  In this scenario, in the limit where $2\pi fr$ is large, $\mathcal{P}$ has effectively negligible contribution to any two-point correlations of the timing residuals due to its highly oscillatory nature. Practically, this amounts to dropping the $P(f;r)$ term in the timing residuals. Here, we consider fully coherent matched-filter maps, where the pulsar term is well constrained. The coherent matched-filter maps then exhibits a distinctive sinusoidal intensity modulation which depends only on the polar angle originating from the pulsar position (i.e. only on $\theta$ for a pulsar located at the north pole).  As we add pulsars, bright islands form in the matched filter map as these sinusoidal patterns add constructively.  These bright fringes allow for gravitational wave localization far beyond the resolution limit derived using the earth term only  (see Fig. \ref{fig:nu30}). However, distinguishing the primary peak and the secondary ``side lobes'' (i.e. bright secondary islands offset from the GW position) depends on the signal to noise ratio. As the dimensionless parameter $\nu$ increases, the fringe spacing gets smaller and smaller and the side lobes peak at higher and higher fractions of the maximum pixel (which corresponds to the location of the GW).

\section{Localizing a single GW}\label{sec:localizing}

%
Our goal in this section is to estimate the number of precisely measured pulsar distances required to reach diffraction-limited resolution for a single source. For a single, circularly polarized source, we define $f_{\rm sky}$ to be the fraction of pixels in the $\rm \operatorname{Re}[\bf L]$ or $\rm \operatorname{Re}[\bf R]$ sky map that exceed a threshold of $\rho = (1-1/\rm SNR) * p_{\rm max}$, for left and right polarized sources, respectively. Here, $\rm p_{\rm max}$ is the maximum pixel value in $\rm \operatorname{Re}[\bf L]$ or $\rm \operatorname{Re}[\bf R]$. We introduce $\bar{f}_{\rm sky} = f_{\rm sky}/\rm SNR$ which characterizes the resolving power of the PTA for a fixed SNR. $\bar{f}_{\rm sky}$ decreases with each additional pulsar distance, $N_{\rm dist}$, as:

\begin{align}
    4 \pi \bar{f}_{\rm sky} \sim (1/\mathrm{SNR})^{N_{\mathrm{dist}}/2},
    \label{eq:scaling_relation}
\end{align}
until reaching a diffraction limited resolution $\delta \theta_{\rm diff}^2 \times \left(\frac{1}{\rm SNR}\right)^2$. At this point, the resolution is no longer improved by adding pulsar distances. Here, we assume equidistant pulsars so that our analysis is based on a single value of $\nu$ and converges to the corresponding resolution. If instead we choose a set of unequal pulsar distances, each distance has its own $\nu$ value,
and the limiting area depends on an average of multiple $\nu$ values. The scaling relation in this case does not change, only the maximum resolution.

The scaling relation in Eq.~\ref{eq:scaling_relation} is found heuristically from the flat-sky limit (see Appendix). This is a relevant limit for coherent PTA analysis as typical values of $\nu$ are quite large ($\gtrsim 100$) corresponding to interference fringes on the order of $\lesssim 1$ arcmin (at this scale, curvature causes fractional distance deviations on the order of $10^{-8}$).  In this limit, a matched-filter map containing a single source, normalized to have a peak value of one, can be represented as $|f(\boldsymbol{x})|^2$ where $f(\boldsymbol{x}) = \frac{1}{N_{\rm dist}} \sum^{N_{\rm dist}}_{i=1} \cos(\boldsymbol{k}_i \cdot \boldsymbol{x})$.  Here $\boldsymbol{x}$ and $\boldsymbol{k}_i$ are two-dimensional vectors on a plane, and the GW source angular position is chosen as the origin. $\boldsymbol{k}_i$ are random wavenumbers with RMS $\sigma_{|k|} \sim r/\lambda_{\rm GW} = \nu$.

\begin{figure}[htbp]
    \centering
    \includegraphics[width=0.8\linewidth]{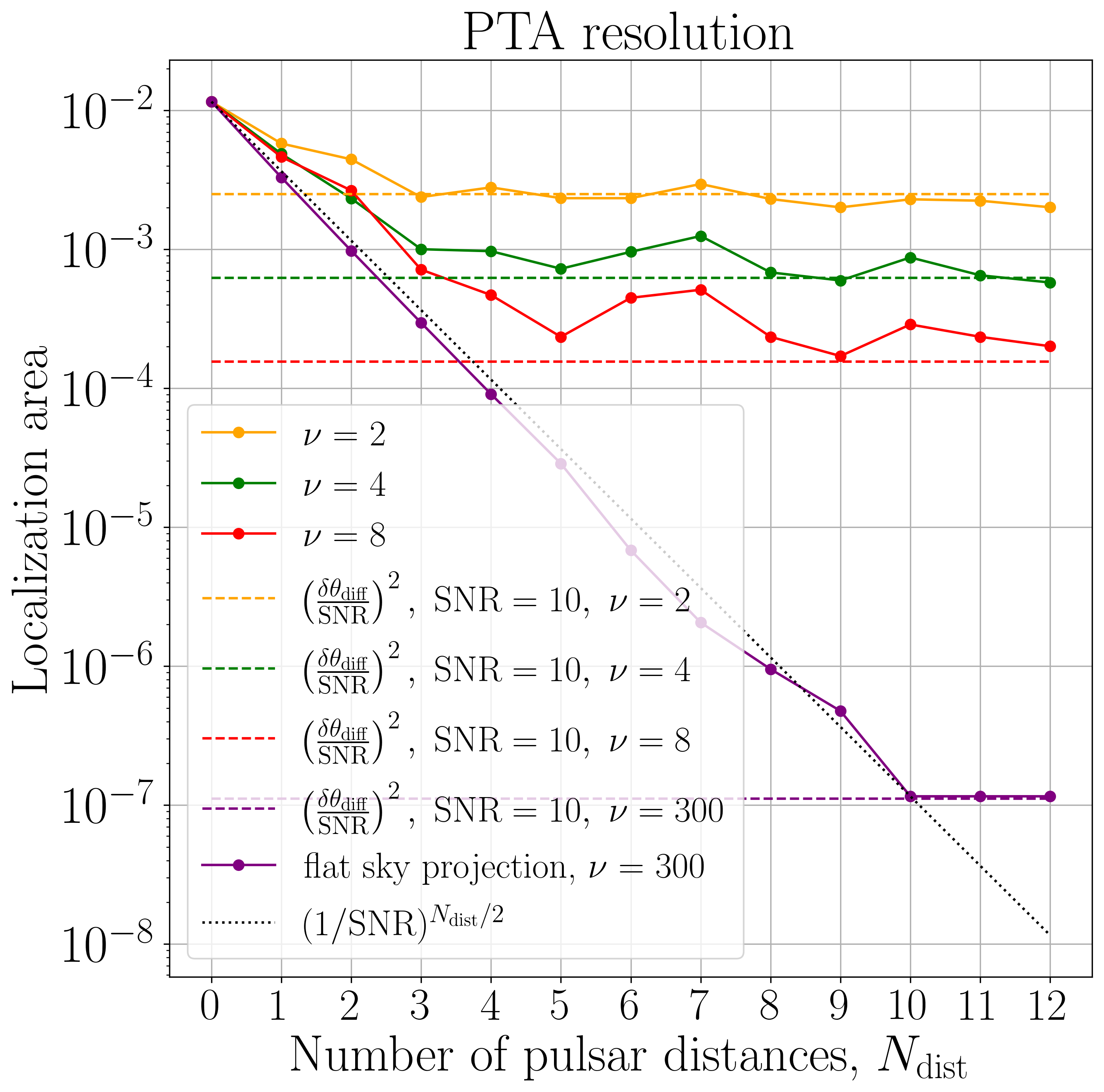}
    \caption{We compare the angular localization area $4 \pi \bar{f}_{\rm sky}$ (solid lines) for increasing numbers of precisely constrained pulsar distances to the diffraction limited resolution, $(\delta \theta_{\rm diff} )^2 \times \left( \frac{1}{\rm SNR}\right)^2$ (dashed lines), for different values of $\nu \equiv r/\lambda_{\rm GW}$. In the small $\nu$ regime ($\nu = 2,4,8$) we conduct a full sky coherent analysis by considering a single, left-circularly polarized GW at a fixed position $\hat{n} = (1,\pi/2,0)$. $N_{\rm dist}$ equidistant pulsars are distributed randomly across the sky.  The resulting matched-filter map, $\operatorname{Re}[L]$, computed for the pulsar term only is then added to an incoherent map (i.e. the matched-filter map, $\operatorname{Re}[L]$, for a highly over-dense PTA, $N_{\rm pulsars} = 4800$ chosen for visualization purposes only, using Earth term only). Adding the coherent maps to the incoherent, Earth-term only map simulates gradually adding individual pulsar distances to a PTA with poorly constrained distances.  The two maps are normalized by their maximum pixel value before being added together. To generate \texttt{HEALPix} maps we choose $N_{\rm side} = 70$ to resolve interference fringes. Considering a more realistic value of $\nu = 300$ approximated from an expected average pulsar distance of $\sim 300$ parsecs for the future $\sim 10$ best constrained pulsar distances, $4 \pi \bar{f}_{\rm sky}$ for $\nu = 300$ (solid purple) is simulated on a small, flat patch of sky. This is achieved by populating a 2D array with $N_{\rm dist}$ randomly placed ones (the array is set to 0 everywhere else). We then take the 2D Fourier transform to get a sum of complex plane waves. The modulus squared of the Fourier transform is the flat sky projection of $\operatorname{Re}[L]$ centered at the GW source. We measure the fraction of the sky above $90 \%$ of the maximum and plot the corresponding $4 \pi \bar{f}_{\rm sky}$. The scaling relation prior to reaching diffraction limited resolution is $(1/\rm SNR)^{N_{\rm dist}/2}$ (black dotted line).}
    \label{fig:Npulsar}
\end{figure}

To demonstrate the full-sky procedure (which is the relevant procedure, given real data) we consider small values of $\nu$, which produce computationally simple-to-resolve fringe patterns. In Fig.~\ref{fig:Npulsar}, we construct $\rm \operatorname{Re}[\bf L]$ maps of single, left handed GWs for small values of $\nu$, which we can easily resolve using a \texttt{HEALPix} pixelation with $N_{\rm side} = 70$ and $N_{\rm pix} = 12 N_{\rm side}^2$ (where $N_{\rm pix}$ is the number of pixels in the map), to support the scaling relation found in the flat-sky regime. We estimate the angular resolution by calculating iso-contours of the threshold value $\rho$. We choose a fiducial SNR of $10$\footnote{In Ref.~\cite{Jow}, we demonstrate that measuring an SNR of $3$ (which is the current sensitivity level on the GWB) using a stochastic, isotropic (Hellings-Downs) analysis on a single source corresponds to a map-space pixel estimator significance of 5.2.  Approximately half the signal strength (the value of the maximum pixel) in a single source map is in the pulsar term (see Fig. \ref{fig:nu30}), meaning a coherent analysis at the same noise level for the same source will have an SNR of roughly 10.6. This sensitivity estimate includes the effects of a look-elsewhere factor of $10^8$ as an estimate of the number of independent pixels in the coherent matched-filter map.
}; i.e. assuming that the individual source can be detected with a SNR of 10, we ask how well that source can be localized. At this sensitivity, angular resolution depends on an iso-contour at $90\%$ of the maximum pixel value.  It is only within this contour that adding random gaussian noise with a standard deviation of $1/10$ of the maximum pixel would have an effect large enough to match or exceed the value of the peak and confuse GW localization. 

Any realistic PTA will have a mixture of pulsars with well-constrained and effectively unknown distances. To reflect this, we start with a dense PTA and consider the pulsar response from the Earth-term. We fix pulsar locations at the center of each pixel of a \texttt{HEALPix} pixelation with $N_{\rm side} = 20$ corresponding to 4800 pulsars. This is significantly more than the $\sim 20$ pulsars required to achieve the maximum angular resolution of this incoherent map ($7^{\circ}$ at an SNR of $10$) \citep{Jow}. This extreme PTA density is used only for visualization purposes to produce smooth maps. A more realistic pulsar density, $\sim 100$ pulsars, produces the same final estimate on number of distances.  To simulate the effects of gradually adding coherent information, we construct matched-filter maps by adding the pulsar-term for an increasing number of pulsars. We normalize the coherent and incoherent maps using the maximum pixel value in each map before adding the two.  We repeat this procedure for a few numerically resolvable values of $\nu$ to validate the scaling relation (Eq.~\ref{eq:scaling_relation}) on a full sphere.

For realistic GW frequencies and pulsar distances, $\nu \gtrsim 100$, resolving interference fringes becomes more computationally challenging due to the large number of spatial pixels in the aperture matrix.  The number of rings in a \texttt{HEALPix} map is proportional to $N_{\rm side}$.  We find that $N_{\rm side}$ must be at least $\sim 8 \nu$ to resolve the polar, sinusoidal pattern. The computation time scales as $\sim N_{\rm side}^2$, resulting in a steep $\sim\nu^4$ scaling in the number of pixels. Fast, full-sky algorithms for rapid computation on a sphere are available \citep{Padmanabhan, Pen} and can be used to build a coherent, matched-filter analysis pipeline on the full sky for real data (we leave this endeavor to future analysis). However, the goal of this work is to make an order-of-magnitude estimate on the number of precise pulsar distance measurements required to reach diffraction-limited resolution.  For this purpose the flat sky approximation is sufficient.  We simulate in this regime by taking a 2D Fourier transform of a sum of $N_{\rm dist}$ delta functions randomly distributed on a 2D plane. This yields a sum of $N_{\rm dist}$ plane waves.  The modulus squared of the Fourier transform approximates the matched-filter map in the region around the GW.  We can use this flat sky projection to calculate $4 \pi \bar{f}_{\rm sky}$ for large values of $\nu$.

Using the results of Fig.~\ref{fig:Npulsar}, we can now answer the question: how many pulsar distances are needed to reach the diffraction-limited resolution? We consider the maximally sensitive PTA observing frequency of $\sim 10^{-8}$ Hz \cite{Baier} (corresponding to a wavelength of $\sim 1$ parsec) and a set of equidistant pulsars at a distance of $300$ parsecs. We estimate we can reach the diffraction-limited resolution for $\nu = 300$, $\delta \theta_{\rm diff} \times \rm \frac{1}{\rm SNR} = \frac{1}{\nu} \times \rm \frac{1}{\rm SNR} \approx 1$ arcmin for $\rm SNR = 10$, with roughly $10$ pulsars. A corollary of this analysis is that, due to the sharp scaling of the resolution with $N_{\rm dist}$, each pulsar distance added to the analysis can improve the resolution by nearly an order of magnitude, until the diffraction limit is reached. We stress, therefore, that a coherent PTA analysis including distance information should not be regarded as a future prospect, but, rather, a method that can be used imminently with even one or two precisely measured pulsar distances to achieve significant gains.


\section{Discussion}



One significant detail we have neglected thus far is the position-dependent sensitivity of GW localization with this method of coherent analysis. Using the aperture matrices (for simplicity we will just consider the left matrix and left polarized GWs), we can construct an $N_{\rm pixel} \times N_{\rm pixel}$ covariance matrix:
\begin{align}
    C^L = {A^L_{ij} (r)}^{\dagger} {A^L_{ij} (r)}.
\end{align}
Matched-filter maps do not necessarily exhibit maxima at the true source positions for general GW locations. The diagonal of $C_L$ is itself an $N_{\rm pixel}$ sky map that measures the localizability of different source positions for the PTA configuration. The location and size of the high and poor localizability regions depends on both the configuration of the pulsars on the sky, as well as their distances, $\nu$.  For a small number of pulsars ($\sim 2$), most GWs cannot be precisely localized; however, as more pulsars are added ``blind-spots'' shrink dramatically and diffraction-limited localizability becomes possible for most GW positions.  Even in blind-spots, the worst resolution is set by the Earth-term only information (i.e. when pulsars are effectively infinitely far away).  Localization in these cases is still good to $0.1 \%$ of the sky (roughly $7^{\circ}$) for an SNR of 10 (or, alternatively, $29^\circ$ for an SNR of 1 \citep{Jow}).

In our analysis, we have assumed that precisely constrained pulsar distances are measured exactly and have effectively no uncertainty. Distance uncertainty may cause a significant angular displacement of the maximum pixel. Distance uncertainties generate phase uncertainties in the pulsar term. In the 1D flat sky limit, $f(x)$ becomes $f(x) = \frac{1}{N_{\rm dist}} \sum^{N_{\rm dist}}_{i = 1} \cos(k_i x - \phi_i)$ where $\phi_i$ is phase induced by uncertainty in the $i^{\rm th}$ pulsar distance.  The angular shift in the location of the maximum in the perturbative limit is $\delta x_{\rm max} = \frac{\sum_i k_i \phi_i}{\sum_i k_i^2}$.  To illustrate the effects of a single distance uncertainty, we consider a simple two-mode example: $f(x) = \cos(x) + \cos(5x -\phi)$. The peak width is set by the largest wavenumber; in this case $\Delta x_{\rm peak} = \pi/5$.  If we consider an uncertainty in the pulsar distance of $10\%$ of the GW wavelength (comparable to the level of uncertainty in PSR J0437-4715), $\phi = 0.1 \times 2 \pi$, and the maximum pixel shifts $\delta x_{\rm max} = 0.12$ radians. This means a distance uncertainty of $10\%$ of the GW wavelength corresponds to a displacement in the peak of $20\%$ of the peak width. In the large-$n$, large-$\nu$ limit, this means $\delta x_{\rm max} \sim 0.2 \times \frac{1}{\sqrt{\rm SNR}} \frac{\lambda_{\rm GW}}{r}$, which even for an SNR of 10 is an effect that is $6\%$ of the diffractive angle. If distance uncertainties are random and independent, the net fractional shift is expected to decrease statistically as $\sim \frac{f}{\sqrt{N_{\rm dist}}}$, where $f$ is the fractional uncertainty in the GW wavelength. However, in practice, distance measurements may exhibit correlated systematics, meaning quantifying effects of distance uncertainties will depend significantly on measurement techniques.

The criterion for a pulsar distance to be ``well constrained'' is that the uncertainty in distance must be less than $\lambda_{\rm GW}/(1+\hat{n} \cdot \hat{r})$ \citep{BoylePen}. In other words, one must be able to measure the phase of the pulsar term to within a radian. It is important to note that this criterion depends on the position of the gravitational wave source itself, $\hat{n}$. Thus, even when the uncertainty of the pulsar distance exceeds a gravitational wavelength, the requirement on the precision becomes less stringent when the source's sky location is close to the pulsar position. This further motivates including distance estimates in current PTA analyses. Even poor distances can be highly constraining for certain GW source localizations. Currently, PSR J0437-4715 has the best constrained pulsar distance ($156.96 \pm 0.11$ parsecs \citep{ReardonBailes}). The next best pulsar distances are for PSR J1909-3744 at $1152 \pm 3$ parsecs \citep{ReadonShannon} from the Shklovskii effect and J0030+0451 at $331 \pm 8$ parsecs \citep{Ding} using VLBI.  We reiterate that the $1$ arcmin resolution we calculate in this paper depends on a short pulsar distance: $\sim 100$ pc ($\nu \sim 100$).  Many pulsars have distances of $\sim 1$ kpc ($\nu \sim 1000$) which corresponds to an order of magnitude improvement in angular resolution. For most pulsars, the most precise distance can be measured from the small, annual pulse arrival difference caused by the motion of the Earth around the sun.  This method yields distances with a precision of $<5\%$, with limited future prospects for improvement \citep{Smits}. In the context of coherent PTA analyses, this will only yield useful phase information for very nearby pulsars, which have less resolving power.

There are several promising methods that stand to further increase the catalogue of pulsars with well-known distances for the purposes of PTAs. Interstellar scintillation produces intensity modulations in frequency and time on signals from compact radio sources from multi-path propagation through the interstellar medium (ISM). Hundreds of images form on discrete screens (often a single screen) along the line-of-sight with separations of $\sim 10$ AU on the screen. These images interfere with each other at the observer and can be leveraged for interstellar interferometry along the scintillating screen \citep{Baker}, achieving angular resolutions $\sim 10^{-2} \rm \mu as$.  In principle, pulsar distances can be measured to as good as $0.4 \%$ in this way \citep{BakerHighres}. Current measurements are not at this level of precision yet \citep{Reardon2024} \citep{Montalvo2025}; however, scintillometry techniques are quickly improving \citep{Mall} \citep{Chen} \citep{Main}. In addition, the annual orbital parallax method uses the apparent motion of the pulsar relative to background quasars to measure pulsar distance. Very long baseline interferometry (VLBI) observations yield precision pulsar localization. VLBI, coupled with annual orbital parallax, is a promising method that has already produced $\sim$\,parsec precision distances. Current VLBI arrays can reach $\sim 20 \mu \rm as$ parallax precision for bright pulsars, a precision which is limited by systematic effects. This corresponds to $\sim$parsec precision distance measurements for pulsars within 500 parsecs. Achieving the much more demanding $\sim 1 \mu \rm as$ precision needed for sub-parsec distances for the general pulsar population will require next-generation facilities. The ngVLA and a DSA-2000 equipped with VLBI outriggers could realistically deliver such performance, while reaching this precision in the southern hemisphere would require building new, wide-field and moderately sensitive VLBI stations. For equatorial pulsars, FAST and GMRT could also help provide the necessary capabilities. Measuring the Shklovskii effect---the apparent change in pulsar spin and orbital period due to radial acceleration---has already achieved precision pulsar distances to sub-parsec precision for PSR J0437 \citep{ReardonBailes} and could potentially add a few high precision measurements for pulsars with long observing times. While we have argued that the diffraction-limited resolution is reached with roughly $10$ pulsars, we stress again that each additional distance measurement will qualitatively improve the angular resolution of a PTA. Thus, even methods with modest prospects for the number of obtainable pulsars distances should be pursued. 

In this work, we have employed a single frequency analysis to calculate resolution.  In reality, we expect at least some SMBHB sources to have eccentric orbits. GWs generated by these systems have power in multiple frequency bins. Far from being a nuisance, this additional information can be used to suppress side lobes by adding matched-filter maps generated at subsequent harmonics of the fundamental map. This analysis will yield an improved angular resolution---although, the effect is small using the fundamental and second harmonic only.

\section{Conclusion}

By constructing coherent matched-filter maps, we arrive at two main conclusions.  First, we show that a PTA can reach the maximum theoretical angular resolution set by the diffraction-limit with only a handful of well-constrained pulsar distances (approximately $10$). While measuring sub-parsec distances for the general millisecond pulsar population requires next-generation facilities, existing techniques are projected to add a few precision distance measurements in the near future. Secondly, due to the steep scaling relationship between the map resolution and the number of constrained distances, every additional high-precision pulsar distance can improve localization by nearly an order of magnitude. This, coupled with the often overlooked subtlety that ``well constrained'' depends on both the gravitational wavelength of interest and the angular proximity of the GW source to a pulsar, imminently motivates a coherent analysis of PTA data and inclusion of all pulsar distance information. Moreover, for a single source, inclusion of the pulsar term can double the detection sensitivity. Thus, we argue that high-resolution matched-filter map making is a promising strategy for distinguishing between a stochastic GWB (of potentially cosmological origin) and discrete astrophysical sources, for which PTAs will likely have sufficient resolving power to characterize and possibly associate with an EM counterpart. Matched-filter mapping is a practical intermediate step in the data processing pipeline extending beyond previous incoherent analyses, which do not take full advantage of pulsar distance information.

\section*{Appendix}

Here, we motivate the origin of the scaling relation in Eq. \ref{eq:scaling_relation}. For large values of $\nu$, the matched filter map $\rm Re[\boldsymbol{L}]$ for a left-handed source (normalized) can be well approximated by $|f(\boldsymbol{x})|^2$ where $f(\boldsymbol{x}) = \frac{1}{N} \sum^{N}_{i=1} \cos(\boldsymbol{k_i} \cdot \boldsymbol{x})$.  Here $\boldsymbol{x}$ and $\boldsymbol{k_i}$ are two dimensional vectors on a flat surface and $N$ is the number of pulsars with known distance.

For simplicity, we will work in 1D and extend our results to 2D at the end. We are interested in answering two questions: what is the limiting width of the central peak as $N \rightarrow \infty$ and what is the probability that at $x \neq 0$ away from the central peak, $|f(x)|^2>\rho$ (where $\rho = (1- 1/\rm SNR)$ as $f(x)$ is normalized to peak at unity).

To answer the first question, we expand $f(x)$ around $x=0$:

\begin{equation}
    f(x) \approx \frac{1}{N} \sum_i(1-\frac{(k_ix)^2}{2}) = 1-\frac{\left< k^2 \right> x^2}{2} 
\end{equation}
where $\left< k^2 \right> = \frac{1}{N} \sum_i k_i^2$ is the mean-square of the wavenumbers. In the actual matched-filter map $\rm Re \left[ \boldsymbol{L} \right]$, the RMS $\sigma_k = \sqrt{\left< k^2 \right>} = 2 \pi r/\lambda_{\rm GW}$. Keeping only the lowest order in $x$, $f(x)^2 \approx 1-\sigma_k^2 x^2$.  We are interested in the width of $f(x)^2 >\rho$ which is obtained by the condition $f(\Delta x/2) = \rho$:

\begin{equation}
    \Delta x = 2 (1-\rho)^{1/2}/\sigma_k \sim \frac{\lambda_{\rm GW}}{r} \rm SNR ^{-1/2}.
\end{equation}

Thus, the main peak width can be written in terms of $\delta \theta_{\rm diff}$ as $\Delta x \sim \delta\theta_{\rm diff} \rm SNR^{-1/2}$.  In 2D, we are interested in a peak area which is simply $(\Delta x)^2 \sim (\delta\theta_{\rm diff})^2 \rm SNR^{-1}$.  This is the origin of the factor of $\rm SNR$ introduced in $\bar{f}_{\rm sky}$.

In our second question, we are interested in the low $N$ regime where side-lobes have some non-zero probability of exceeding the threshold $\rho$.  To see how this probability scales with $N$, we consider $x$ far away from the central peak.  In this regime, $f(x) = \frac{1}{N} \Sigma_{i} \cos(\theta_i)$ where $\theta_i$ is a uniform random angle between $[0,2\pi)$.  We want to know the probability that 
$f(\mathbf{x})^2 > \rho$ or equivalently, $f(\mathbf{x}) > \sqrt{\rho} = \sqrt{1- 1/\mathrm{SNR}}$. For high $\rm SNR$ ($\rho$ close to one), this will only occur if all $\theta_i$ are sufficiently aligned close to zero.  We approximate the probability of the sum exceeding the threshold by requiring that every $\cos(\theta_i) > \sqrt{1- 1/\mathrm{SNR}}$.  This occurs when:

\begin{equation}
    \theta_i \lesssim \rm SNR^{-1/4},
    \label{eq:one_cosine}
\end{equation}
where we have used the small angle approximation of cosine and kept only powers of $\rm SNR$.  The probability for all N cosine functions to meet this condition is the probability of one raised to the N:

\begin{equation}
    P(|f(x)|^2 > \rho) \sim \rm SNR^{-N/4}.
\end{equation}

The last step is to extend this relation to 2D where Eq.~\ref{eq:one_cosine} becomes $\theta_i < \rm SNR^{-1/2}$, as the probability $P(|\boldsymbol{\theta}|<\epsilon)$ goes as $\epsilon^2$ when $\boldsymbol{\theta}$ is a 2D vector.  With this extension, we recover our scaling relation:

\begin{equation}
    P(|f(\boldsymbol{x})|^2 > \rho) \sim \rm SNR^{-N/2}
\end{equation}

\section*{Supplemental Material}
An animation illustrating the reconstructed sky map procedure used to generate Fig.~\protect\ref{fig:Npulsar} is provided as ancillary material.

\section*{Acknowledgments}

We thank Adam Deller, Stephen Taylor, 
Marc Kamionkowski, Maya Fishbach, Daniel Reardon, and Yacine Ali-Haïmoud for helpful discussions. U.P. is supported by the Natural Sciences and Engineering Research Council of Canada (NSERC) [funding reference number RGPIN-2019-06770, ALLRP 586559-23, RGPIN-2025-06396] , Canadian Institute for Advanced Research (CIFAR), Ontario Research Fund (ORF-RE Fund), and AMD AI Quantum Astro.

\bibliography{references}

\end{document}